\documentclass[journal=jacsat,manuscript=article]{achemso}

\usepackage[version=3]{mhchem} 
\usepackage{amsmath}
\usepackage{amssymb}
\usepackage{braket}

\usepackage[defaultcolor=magenta]{changes} 
\setaddedmarkup{\color{blue}#1}

\newcommand{\be}{\begin{equation}}
\newcommand{\ee}{\end{equation}}

\newcommand{\fig}[1]{Fig.~\ref{#1}}

\newcommand{\Fig}[1]{Figure~\ref{#1}}

\newcommand{\eq}[1]{Eq.~\eqref{#1}}

\newcommand{\Eq}[1]{Equation~\eqref{#1}}

\makeatletter
\renewcommand*{\acs@author@fnsymbol}[1]{%
  \ifnum#1=\z@ *
  \else\textsuperscript{\@alph{#1}}
  \fi
}
\makeatother

\author{Ignacio Gustin}
\affiliation{Department of Chemistry, University of Rochester, Rochester, New York 14627, USA}
\author{Chang Woo Kim}
\affiliation{Department of Chemistry, Chonnam National University, Gwangju 61186, South Korea}
\alsoaffiliation{The Research Institute for Molecular Science, Chonnam National University, Gwangju 61186, South Korea}
\author{Ignacio Franco}
\affiliation{Department of Chemistry, University of Rochester, Rochester, New York 14627, USA}
\alsoaffiliation{Department of Physics and Astronomy, University of Rochester, Rochester, New York 14627, USA}
\alsoaffiliation{Institute of Optics, University of Rochester, Rochester, New York 14627 USA}
\email{ignacio.franco@rochester.edu}

\title{Dissipation Pathways in a Photosynthetic Complex}

\abbreviations{IR,NMR,UV}
\keywords{American Chemical Society, \LaTeX}

\begin{document}

\begin{tocentry}
\centering
\includegraphics[width=0.88\textwidth]{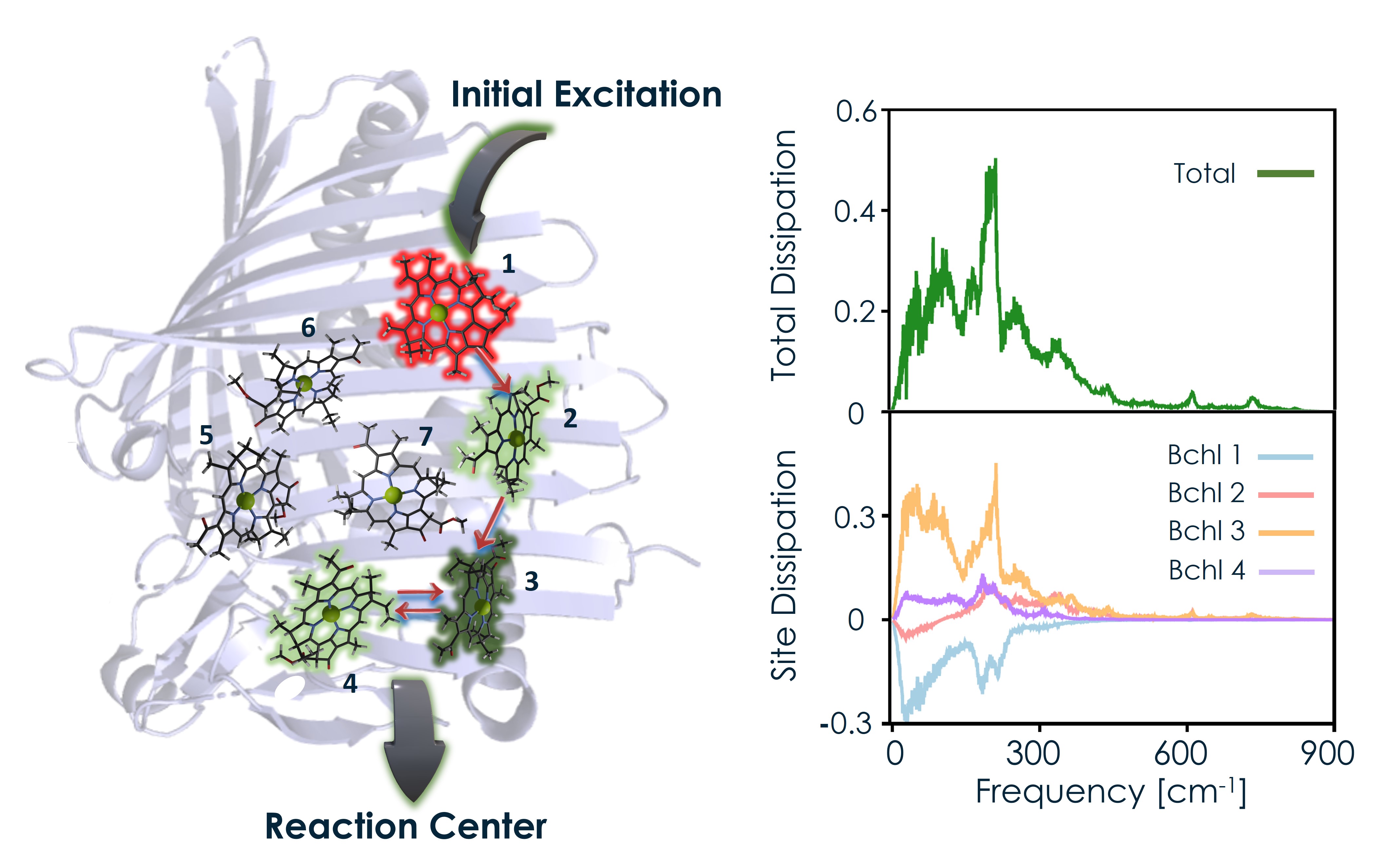}

\end{tocentry}


\begin{abstract}
Determining how energy flows within and between molecules is crucial for understanding chemical reactions, material properties, and even vital processes such as photosynthesis. While the general principles of energy transfer are well established, elucidating the specific molecular pathways by which energy is funneled remains challenging as it requires tracking energy flow in complex molecular environments. Here, we demonstrate how photon excitation energy is partially dissipated in the light-harvesting Fenna-Matthews-Olson (FMO) complex, mediating the excitation energy transfer from light-harvesting chlorosomes to the photosynthetic reaction center in green sulfur bacteria. Specifically, we isolate the contribution of the protein and specific vibrational modes of the pigment molecules to the energy dynamics. For this, we introduce an efficient computational implementation of a recently proposed theory of dissipation pathways for open quantum systems, based on second-order perturbation theory in the electronic couplings. Using it and a state-of-the-art FMO model with highly structured and chromophore-specific spectral densities, we demonstrate that energy dissipation is dominated by low-frequency modes ($<$ 800 cm$^{-1}$) as their energy range is near-resonance with the energy gaps between electronic states of the pigments. We identify the most important mode for dissipation to be in-plane breathing modes ($\sim$200 cm$^{-1}$) of the bacteriochlorophylls in the complex. Conversely, far-detuned intramolecular vibrations with higher frequencies ($>$ 800 cm$^{-1}$) play no role in dissipation. Interestingly, the FMO complex first needs to borrow energy from the environment to release excess photonic energy, indicating that the energy exchange between the system and the thermal environment is not strictly unidirectional in time but involves a transient thermally activated step. Beyond their fundamental value, these insights can guide the development of artificial light-harvesting devices and, more broadly, engineer environments for chemical and quantum control tasks.
\end{abstract}


Photosynthesis is a critically important light-induced process that enables plants, algae, and bacteria to convert solar energy into biochemical fuel with remarkable efficiency.\cite{blankenship2021molecular,baker2008chlorophyll} This transformation begins with the creation of excitons by solar photons, which travel through a number of pigment molecules until reaching the reaction center, where the energy is transformed into chemical fuel.\cite{mirkovic2017light}  During this complex energy-transfer process, excitons dissipate excess energy into the nuclear environment. Although the fundamental principles of energy transfer are well established, mapping the precise molecular pathways of this energy flow remains a formidable task as it requires tracking energy dynamics in complex molecular environments.

Specifically, it is not understood \textit{how} individual components of the photosynthetic system, including the pigment molecules and their surrounding protein environment, contribute to the overall energy dissipation. Understanding these dissipation pathways is essential for characterizing the energy flow and overall dynamics of photosynthetic processes,\cite{mirkovic2017light,jang2018delocalized,cao2020quantum} and is needed to address fundamental questions such as: How do the different components of the pigment molecules contribute to the overall dissipation? Are there specific vibrations that contribute the most to energy dissipation? Is energy transfer a unidirectional process, or do pigment molecules both absorb from and dissipate energy into the environment? Addressing these basic questions requires general strategies to connect chemical structure to energy transfer dynamics in complex molecular systems. 

From an experimental perspective, significant spectroscopic efforts have been made to understand the dynamics of energy transfer within photosynthetic complexes.\cite{allodi2018redox,brixner2005two,engel2007evidence,panitchayangkoon2010long,thyrhaug2016exciton,dostal2016situ,duan2017nature,maiuri2018coherent,thyrhaug2018identification,higgins2021photosynthesis,segatta_quantum_2017} 
However, it remains challenging to unravel the intricate energy transfer dynamics in complex chemical environments because the spectral congestion\cite{arsenault2020role,lambrev2020insights} often prevents disentangling contributions by individual components of the pigments and the protein. Progress has been made through two-dimensional electronic spectroscopy (2DES)\cite{biswas2022coherent}, which allows tracking exciton dynamics along two frequency axes with femtosecond resolution. Nevertheless, even with these advanced techniques, quantifying the contribution of individual nuclear modes to the dissipation remains elusive. This is because the electronic states couple to numerous vibrations, and crucially, 2DES projects this high-dimensional vibronic space onto a reduced measurement dimension, thereby limiting the resolution of individual nuclear contributions to energy dissipation.

\begin{figure*}[htpb]
   \centering
\includegraphics[width=1\textwidth]{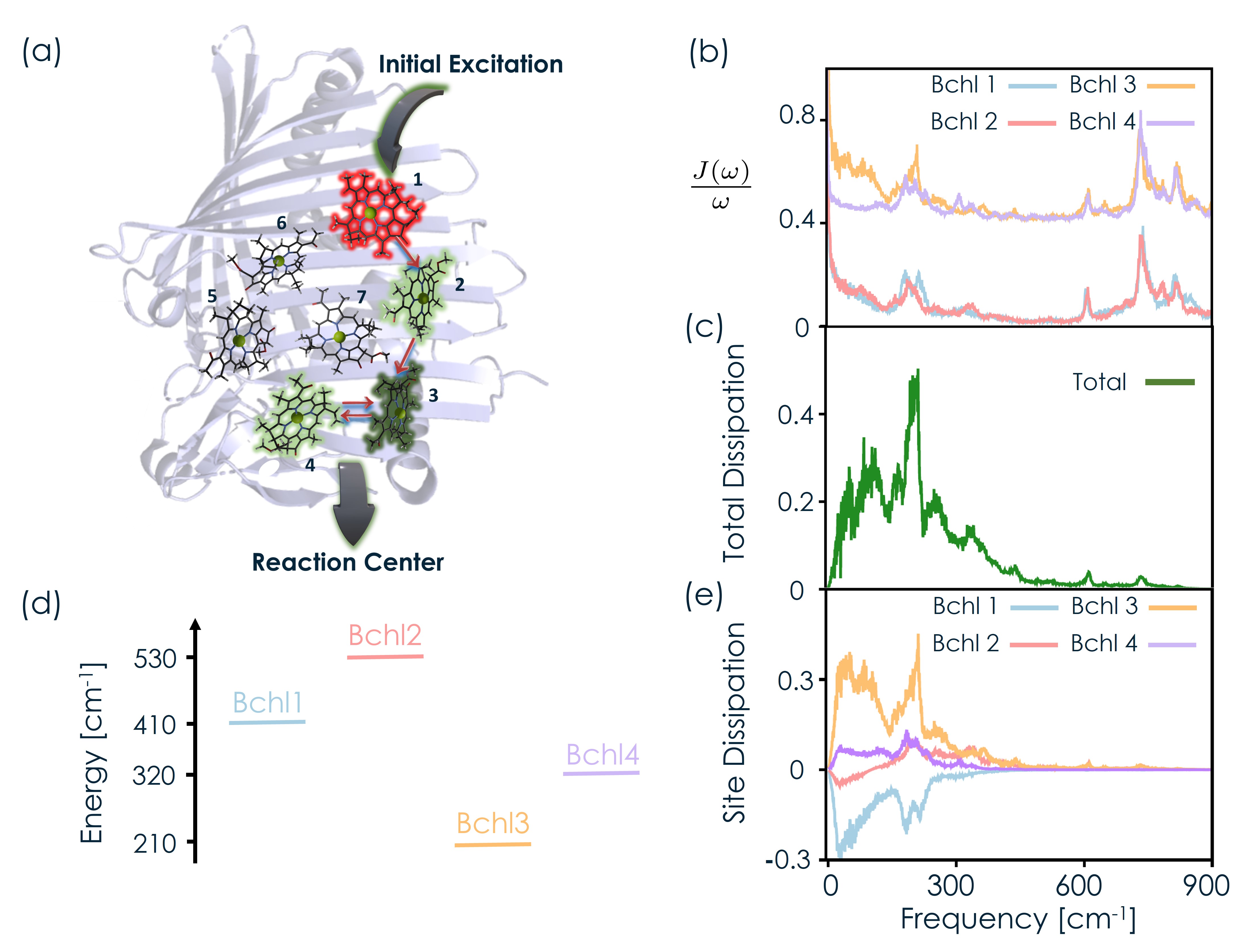}
\caption{\textbf{Dissipation pathways in the FMO complex.}  (a) Schematic illustration of the energy dynamics in the FMO complex. The Bacteriochlorophyll (Bchl) chromophores are labeled with bold numbers, and the intensity of the green (red) color indicates the amount of energy being dissipated (absorbed). (b) Scaled spectral density $J(\omega)/\omega$ for Bchl1 (light blue), Bchl2 (pink), Bchl3 (orange), and Bchl4 (purple). The baselines for Bchl3 and Bchl4 were vertically shifted for visual clarity. (c) Total dissipation as a function of the environmental frequency. (d) Site energies for Bchl1, Bchl2, Bchl3 and Bchl4. (e) Dissipation decomposition in terms of individual Bchl chromophores.} 
\label{fig:FMO-scheme}
\end{figure*}

From a theory perspective, the task requires a method to decompose energy dissipation into contributions of the protein and specific vibrational modes of the pigment molecules.  The challenge is that this requires quantum dynamical information about this complex chemical
environment, and the computational cost of obtaining this information remains beyond the reach of state-of-the-art methods in quantum dynamics. For instance, explicit approaches such as the Multi-configuration time-dependent Hartree (MCTDH)\cite{beck2000multiconfiguration} accurately model environmental dynamics through direct wavefunction propagation, yet they become computationally intractable for macroscopic chemical environments. Trajectory-based mixed quantum-classical methods can also treat complex chemical environments.\cite{sokolov2024non, kim2012all} However, they suffer from artifacts like zero-point energy leakage and negative populations that lead to unphysical dissipation computations.\cite{kim2014improving,kim2020toward} Moreover, their classical treatment of the environment does not correctly capture important quantum processes such as spontaneous emission. By contrast, quantum master equations and related techniques\cite{kundu_tight_2022,tempo,varvelo2021formally,Bose2022multisite} can treat complex chemical environments implicitly by focusing on the effect in the system's dynamics. This scalability, however, comes at the cost of discarding information about the environment's quantum dynamics.

In response to this challenge, we recently developed\cite{kim2024general1,kim2021theory} a general fully quantum theory to quantify and resolve dissipation pathways of quantum systems immersed in highly structured quantum thermal environments, including both harmonic and anharmonic baths. The theory, based on second-order perturbation theory of the off-diagonal system’s couplings, captures the energy dissipated into individual environmental modes while avoiding explicitly propagating the quantum dynamics of the environment, thereby opening a systematic and computationally tractable path to investigate the energy flow between the system and its environment. Here, we introduce an efficient computational implementation (see Theoretical Methods) of this theory that now enables access to the dissipation pathways of complex molecular systems while ensuring that the computational cost remains tractable. 

We use this strategy to elucidate how photon excitation energy is dissipated in the Fenna-Matthews-Olson (FMO) complex\cite{fenna1975chlorophyll}. This complex mediates the transfer of excitation energy from light-harvesting chlorosomes to the photosynthetic reaction center in green sulfur bacteria.\cite{blankenship2021molecular,jang2018delocalized} While the structure and exciton dynamics in the FMO have been well studied\cite{adolphs2006proteins,gonzalez-soria_parametric_2020,kell_effect_2016,christensson_origin_2012,gillis_theoretical_2015,rivera_influence_2013,wendling_electronvibrational_2000,lee_modeling_2016,brixner2005two,engel2007evidence,kim2018excited,higgins2021photosynthesis,orf_evidence_2016,maly_role_2016,abramavicius_excitation_2014,padula_chromophore-dependent_2017,schulze_explicit_2015,olbrich_theory_2011,moix2011efficient,chandrasekaran2015influence,milder2010revisiting,tronrud2009structural,schmidt2011eighth,camara2003structure,olbrich2011atomistic,chen2022simulation}, the dissipation pathways in the complex have remained elusive to both theory and experiment. Specifically, we isolate the contributions of the protein and intramolecular vibrational modes of the pigment molecules to the energy dynamics. We utilize highly structured and pigment-specific spectral densities to characterize the frequencies of protein and intramolecular vibrations, as well as their coupling strengths to electronic excitations in the pigments.\cite{kim2018excited} Since these highly structured spectral densities represent high-dimensional environments with $20+$ vibrational features per pigment, computing the dissipation pathways is beyond the reach of any other state-of-the-art method in quantum dynamics. 

As discussed below, vibrations with frequencies $<$ 800 cm$^{-1}$ dominate the energy transfer dynamics, with in-plane breathing modes of the bacteriochlorophylls being the most important ones. While the importance of low-frequency vibrations in the FMO complex has been recognized\cite{nalbach2015vibronically} our work moves beyond this, elucidating how and why specific modes govern the dissipation dynamics with no a priori assumptions. Interestingly, the simulations also show that the FMO complex dissipates excess photonic energy through a non-monotonic process, as it initially draws energy from the thermal environment before onsetting the overall dissipative process.  

The introduced strategy provides a path to disentangle the contribution of individual molecules and nuclear vibrations to the energy dynamics during photosynthetic events.\cite{mirkovic2017light,jang2018delocalized,cao2020quantum} This is important for establishing the connection between molecular structure and energy dynamics. Beyond its fundamental value, this connection can be helpful for the design of artificial light-harvesting systems,\cite{hart2021engineering,wasielewski2009self,tang2024simulating} and, more broadly, to engineer environments for chemical and quantum control tasks.\cite{Kienzler2014,Hsu2017,CamposGonzalezAngulo2019,Ng2020,gustin_decoherence_2025,hu2022tuning} 

Our efforts augment and complement previous efforts to investigate energy pathways in molecular arrays. In particular, there are important strategies to track the energy flow in proteins through vibrational energy transfers using atomistic\cite{sagnella2000time,bu2003simulating} and coarse-grained\cite{ishikura2015energy,xu2014vibrational,xu2014communication,leitner2015vibrational,leitner2009frequency,agbo2015vibrational} classical molecular dynamics simulations.  Our approach provides a fully quantum perspective that focuses on the quantification of the dissipation of electronic excitation into nuclear modes. Our efforts also complement strategies\cite{yang2017identifying,pereverzev2009energy} developed to isolate the most important coordinate in complex electron transfer processes using electronic structure methods. Such models can be incorporated into our theoretical framework, enabling the analysis of their associated dissipation pathways.

We analyze how energy is dissipated into the thermal environment in the Fenna-Matthews-Olson (FMO) complex, essential for efficient light harvesting in green sulfur bacteria such as \textit{C. tepidum} and \textit{P. aestuarii}. The FMO complex acts as a molecular bridge, facilitating the transfer of excitation energy from the chlorosome, where photons are absorbed, to the reaction center, where this energy is converted into chemical fuel. While this transfer is mediated by exciton dynamics within a trimeric FMO complex, research suggests that each exciton's pathway is confined to a single monomer during transfer\cite{jang2018delocalized}, each housing eight bacteriochlorophyll (Bchl) molecules (see \fig{fig:FMO-scheme}a). Therefore, we focus on a single monomer. 

To describe the energies and couplings of the excited electronic states in the Bchl array, we adopt the Adolphs and Renger\cite{adolphs2006proteins} Hamiltonian (see Sec. SI in the Supplementary Information). The model includes 7 of the 8 BChl as one of them acts as an incoherent energy funnel exlusively connected to BChl1.\cite{moix2011efficient}
While reported site energies can differ by up to 100 cm$^{-1}$ across studies\cite{higashi2016quantitative}, the overall energy landscape and transfer pathways remain consistent across different parameter sets. The Adolphs and Ranger parameterization provides a well-established model that captures this consistent landscape and accurately reproduces the optical spectra of the FMO complex.\cite{adolphs2006proteins} The interaction between the Bchls and the surrounding environment is described by Bchl-specific spectral densities, computed by Kim and coworkers\cite{kim2018excited}, which accurately capture the nuclear environment's vibrational frequencies ($\omega$) and their coupling strengths with the Bchls electronic excited state with a frequency resolution of $10^{-4}$ cm$^{-1}$. These spectral densities show good agreement with available experiments and consist of broad features at frequencies below $200$ cm$^{-1}$ that account for protein contributions and a series of sharp peaks to describe the influence of intramolecular vibrational modes. In \fig{fig:FMO-scheme}b, we detail a segment of the spectral density scaled by the environment frequency, $J(\omega)/\omega$, for the chromophores that are most important in the process of absorption/dissipation of energy. The remaining spectral densities are presented in Fig. S1. The reliability of our spectral densities, particularly in the challenging low-frequency region, is ensured by the extensive 100 ns MD trajectory employed in its construction. This approach provides a high frequency resolution of 3 $\times$ 10$^{-4}$ cm$^{-1}$.\cite{kim2018excited}

Our analysis is based on a recently proposed fully quantum theory of dissipation pathways\cite{kim2024general1,kim2024general2}, see Theoretical Methods for details, which is used to quantify the amount of energy channeled from the Bchls in the FMO into the complex's vibrational modes, including both protein and intramolecular contributions. The applicability of our theory, which assumes weak electronic coupling, to compute dissipation pathways in the FMO complex is validated against the numerically exact hierarchical equation of motion (HEOM)\cite{tanimura2020numerically, Ikeda2020,kim_extracting_2022} using illustrative models (see Secs. SII-SIII in the Supplementary Information), showing semi-quantitative agreement. We note that direct HEOM calculations of dissipation pathways incorporating the highly structured spectral densities of the FMO complex are currently computationally prohibitive. 
Thus, our benchmark, using representative models that capture the essential features of the FMO, represents the state of the art.

\Fig{fig:FMO-scheme} shows the contributions of specific environmental modes at frequency $\omega$ to the overall asymptotic energy absorption/dissipation in the overall complex and in each chromophore when the initial excitation is placed in Bchl1, the site closest to the chlorosome.\cite{wen2009membrane} This electronic excitation moves through the complex until it reaches Bchl3 and Bchl4, the chromophores closest to the reaction center.\cite{adolphs2006proteins} In \fig{fig:FMO-scheme}a, chromophores are labeled with bold numbers, and the intensity of the green (or red) color indicates the relative amount of energy dissipated (or absorbed) at each chromophore in the overall process. 

\Fig{fig:FMO-scheme}c shows the overall frequency-resolved dissipation profile, which reveals that the dissipation of energy is driven by the low-frequency modes ($<$ 800 cm$^{-1}$). This is because their energy is near-resonant with the energy gaps between excited electronic states in the complex, see \fig{fig:FMO-scheme}d. In particular, we identify environmental modes around 200 cm$^{-1}$ as the most important modes for dissipation in the FMO complex, giving the most important contribution in \fig{fig:FMO-scheme}c. Remarkably, despite being present in the spectral densities\cite{kim2018excited}, see Fig. S1, high-frequency vibrational modes ($>$ 800 cm$^{-1}$) do not contribute to the energy dissipation in this complex as their energy is far detuned from the energy gaps between excited electronic states. This is consistent with previous analyses in the FMO complex\cite{abramavicius_excitation_2014,maly_role_2016,olbrich_theory_2011,schulze_explicit_2015,padula_chromophore-dependent_2017,wu_efficient_2012}  that showed that high-frequency components are not necessary to capture exciton dynamics accurately.

In \fig{fig:FMO-scheme}e, we decompose the overall dissipation into the contribution of the nuclear modes coupled to the most relevant Bchls (1-4). Interestingly, Bchl1 (light blue) has a ``negative" dissipation, which means it absorbs energy from the environment. By contrast, Bchl2 (pink), Bchl3 (orange), and Bchl4 (purple) release energy into the environment, with Bchl3 making the most significant contribution. The remaining Bchls are not shown as they have a minor impact on the exciton population and energy dissipation dynamics. 

\begin{figure}[htb]
    \centering
\includegraphics[width=0.4\textwidth]{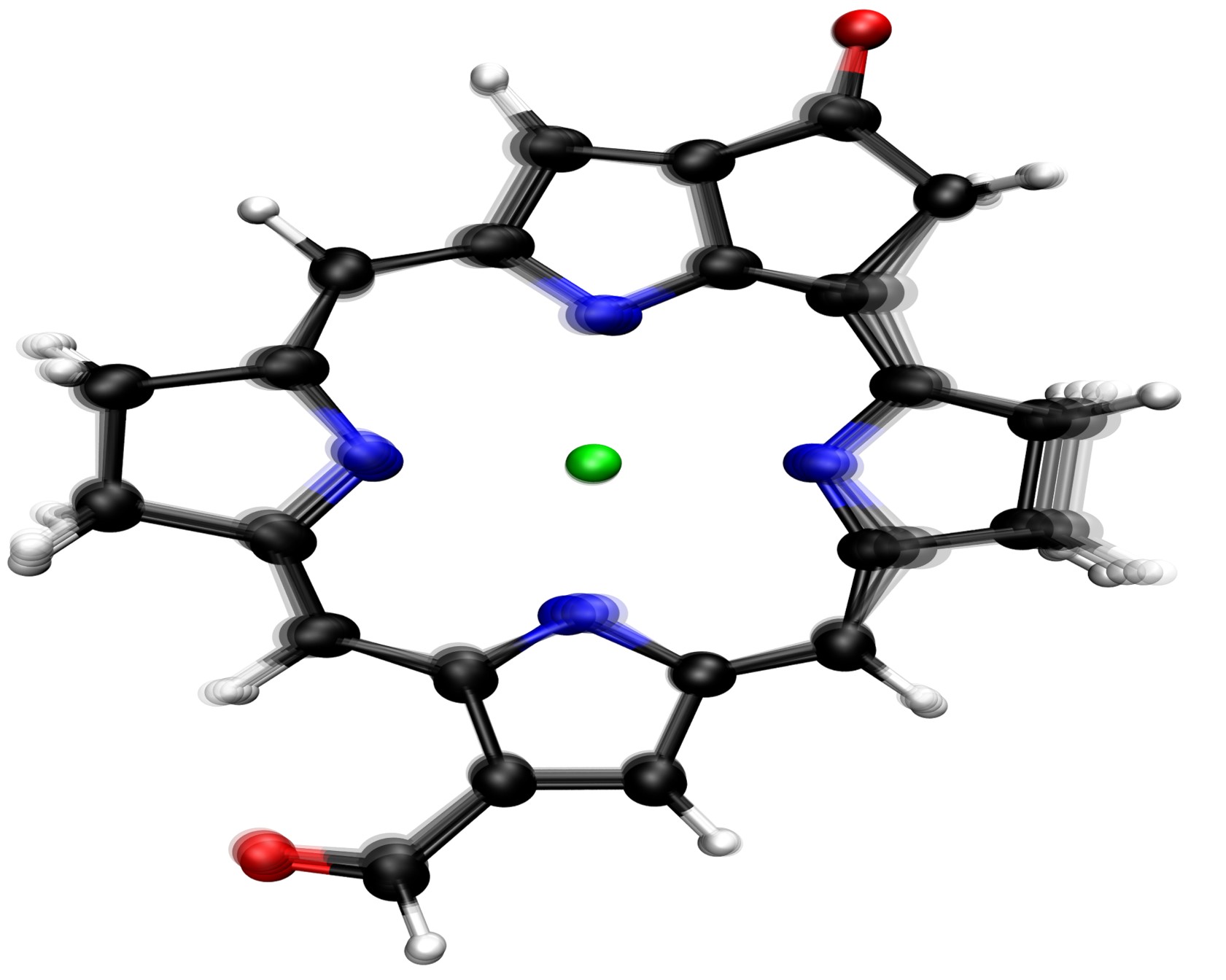}
    \caption{\textbf{Vibrational mode driving energy dissipation in the FMO complex.} The figure shows the vibrational mode around 200 cm$^{-1}$ in Bchl3, crucial for energy dissipation. This mode is remarkably similar across all chromophores (see animations in the Supplementary Materials). The mode was extracted from molecular dynamics simulations as detailed in the Supplementary Information.}
    \label{fig:FMO-NM}
\end{figure}

To elucidate the vibrational modes around 200 cm$^{-1}$ that predominantly contribute to dissipation in the FMO complex across all chromophores, we developed a computational strategy to extract the representative vibrational motion from the molecular dynamics trajectories used for constructing the spectral densities (see Sec. SIV in the Supplementary Information for details). \Fig{fig:FMO-NM} shows the isolated vibrational mode around 200 cm$^{-1}$ for Bchl3, displaying only the Bchl core for enhanced clarity. Interestingly, this vibrational mode, most responsible for dissipation, exhibits a breathing motion pattern and is confined to the plane with no out-of-plane components. We do not show other chromophores as the modes around 200 cm$^{-1}$ exhibit remarkable similarity across all chromophores, including Bchl1, which absorbs energy from the environment. A set of animations showing the vibrational mode for every Bchl around 200 cm$^{-1}$ are available in the supplementary materials. Importantly, the relevance of this vibrational frequency around 200 cm$^{-1}$ has been previously noted in 2DES experiments in the FMO complex where including this vibrational mode in the spectral density was essential for obtaining energy transfer rates that align with experimental observations.\cite{higgins2021photosynthesis} Furthermore, theoretical investigations\cite{nalbach2015vibronically}have also found it necessary to include this mode on an ad hoc basis to reproduce experimental results, reinforcing the consensus of its critical role. By contrast, our work does not presuppose its importance. Instead, the dominant role of the 200 cm$^{-1}$ vibration is a direct outcome of the system's dynamics, providing a first-principles validation of its function and revealing the underlying molecular mechanism driving the dissipation.

To ensure that these findings are not dependent on the initial excitation site, we conducted an identical simulation with the excitation initiated on BChl6. These calculations, detailed in Sec. SV in the Supplementary Information, yield a consistent picture of the energy dissipation dynamics. The process remains dominated by low-frequency nuclear modes ($<$ 800 cm$^{-1}$), with the vibrational mode near 200 cm$^{-1}$ consistently emerging as the most significant contributor to energy relaxation. This confirms that the identified dissipation channel is a robust feature of the system's dynamics.

\begin{figure}[hbtp]
    \centering    \includegraphics[width=0.5\textwidth]{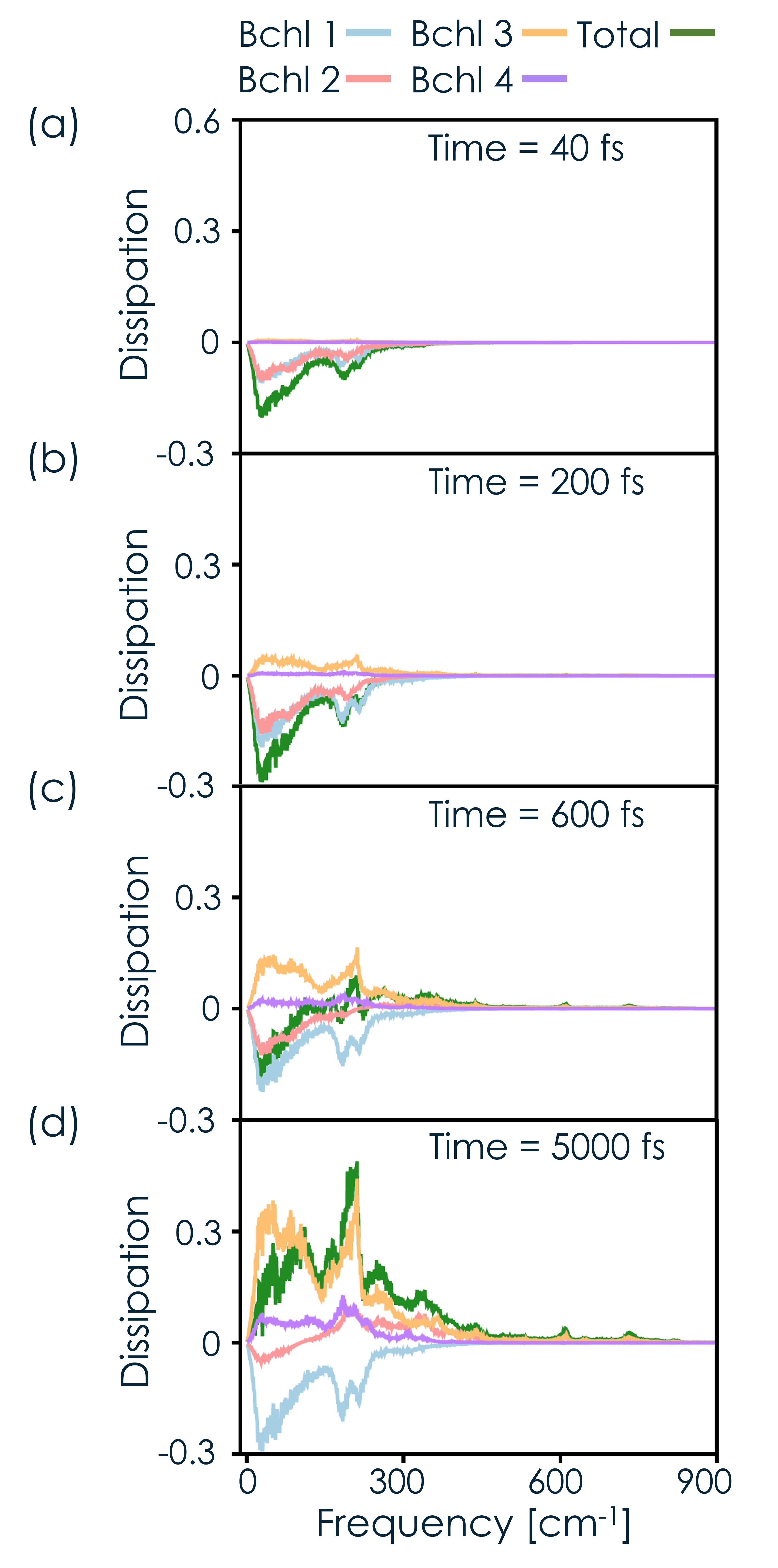}
    \caption{\textbf{Time-dependent energy dissipation in the FMO complex.} The panels show the cumulative dissipation by environmental modes at different frequencies at selected times during the dynamics: (a) 40 fs, (b) 200 fs, (c) 600 fs, and the overall dissipation at (d) 5000 fs.  The colored lines quantify the contributions of nuclear modes belonging to Bchl1 (light blue), Bchl2 (pink), Bchl3 (orange), and Bchl4 (purple). The green line quantifies the total dissipation. Note that negative values of the dissipation indicate energy absorption.}
    \label{fig:FMO-TD}
\end{figure}

We now go beyond the overall dissipation, and scrutinize the role of each Bchl during the energy transfer dynamics.  \fig{fig:FMO-TD} shows the cumulative dissipation over time for the four most significant chromophores and for the overall complex. \fig{fig:FMO-TD}a shows that, initially, the dissipation is negative, indicating that the system is absorbing energy from the environment to facilitate uphill energy transfer from Bchl1 to Bchl2 (see energy level diagram in \fig{fig:FMO-TD}d). Interestingly, both Bchl1 and Bchl2 collectively absorb energy during this phase. This initial energy absorption is consistent with the uphill energy transfer signals involving Bchl1 and Bchl2 observed experimentally in two-dimensional electronic spectroscopy (2DES) experiments measured at 40 femtoseconds (fs).\cite{higgins2021redox}

At 200 fs, as shown in \fig{fig:FMO-TD}b, Bchl3 plays a significant role in overall dissipation due to increased population transfer to this chromophore. By 600 fs, \fig{fig:FMO-TD}c, Bchl1 continues to absorb energy from the environment, while Bchl3 maintains its role in dissipating energy. Meanwhile, Bchl2 reverses its initial energy directionality to dissipate energy into the environment. This change in Bchl2 from absorption to dissipation, along with the ongoing energy dissipation of Bchl3, changes the overall energy profile of the system towards net dissipation into the environment. By 5000 fs, the system reaches a steady state, and it is clear that overall energy is dissipated into the environment across all frequencies. Interestingly, Bchl1 consistently absorbs energy throughout the process. 

This analysis shows that energy flow within the FMO complex is not unidirectional. Some chromophores absorb energy from or dissipate energy into the environment throughout the dynamics, while others can absorb or dissipate energy at different time scales. Furthermore, the analysis shows that not all vibrational modes are equally important for the energy transfer dynamics in the FMO complex. In fact, we identified the in-plane breathing modes around 200 cm$^{-1}$ as the most important for energy transfer.

In conclusion, for the first time, we have successfully isolated the dissipation pathways in the Fenna-Matthews-Olson complex by quantifying contributions by individual environmental components to the energy dynamics.  This was achieved through an efficient computational implementation of our recently developed quantum theory of dissipation pathways\cite{kim2024general1,kim2024general2} combined with a state-of-the-art FMO model incorporating highly structured and Bchl-specific spectral densities. 

Our analysis revealed that energy dissipation in the FMO complex is predominantly driven by low-frequency vibrational modes ($<$ 800 cm$^{-1}$), with in-plane breathing modes around 200 cm$^{-1}$ being the most dominant. High-frequency modes ($>$ 800 cm$^{-1}$) play a negligible role despite their presence in the spectral densities as they are far detuned from the excited state electronic energy gaps in the FMO complex. Furthermore, we demonstrated that energy flow within the FMO complex is non-monotonic, as it involves a transient back-and-forth energy exchange, reflecting a thermally activated step prior to overall dissipation.  Specifically, to release excess photonic energy, the FMO complex must first absorb energy from the environment. We identify BChl1, closest to the chlorosome, as the primary mediator of initial energy absorption and BChl3, nearest to the reaction center, as the primary mediator of energy dissipation.

The ability to precisely map dissipation pathways in complex chemical systems, as demonstrated in this study of the FMO complex, opens new avenues for investigating the role of the environment in energy transfer dynamics.  We envision future applications of this methodology extending beyond the FMO complex to other light-harvesting systems, including, but not limited to, light-harvesting complex II (LH2)\cite{mirkovic2017light}, phycobiliprotein PC645\cite{blau2018local}, cyanobacterial light-harvesting proteins\cite{womick2011vibronic} and other molecular systems\cite{lee2017first,novoderezhkin2004energy,gustin2023mapping,ratsep2007electron,wiethorn2023beyond}. Furthermore, our strategy can aid in guiding emerging efforts to develop more efficient artificial light-harvesting systems\cite{hase2020designing} by identifying the key vibrational modes and pigment molecules involved in energy transfer in natural light-harvesting systems and, more broadly, to advance efforts to control quantum systems via environment engineering.

\section{Theoretical Methods} \label{sec:theory}

In Ref.  \cite{kim2024general1} we developed MQME-D, a general framework for constructing practical and accurate
schemes to isolate dissipation pathways in open quantum systems targeting Markovian Quantum Master Equations (MQME). The formulation incorporates a specific bath degree of freedom into the subsystem component and calculates the change in its energy with Nakajima-Zwanzig projection operator technique. By perturbatively expanding the Liouville-von Neumann equation, the MQME-D elucidates the rate of dissipation expressed
by using traces of operator products, which can be applied for general types of bath and subsystem-bath interaction. The MQME-D also rigorously satisfies  thermodynamic principles such as energy conservation and detailed balance.

Here, we outline the main ideas of our quantum theory of dissipation pathways based on Fermi’s Golden Rule Markovian quantum master equations describing the open quantum dynamics. We focus on the case when the environment consists of a local set of effective harmonic modes, as is the case with photosynthetic complexes. However, the approach can be applied to any harmonic or anharmonic environment, provided that it can be modeled with independent environment degrees of freedom.\cite{kim2024general1,kim2021theory} 

We consider an array of $N$ chromophores that are coupled to one another and to local thermal environments. The state $\ket{A}$ describes a physical situation in which the chromophore $\ket{A}$ is in the excited electronic state while the other chromophores in the complex are in the electronic ground state. The population dynamics of a chromophore $\ket{A}$ is given by
\begin{equation} \label{eq:population}
    \dot{P}_{A}(t) = \sum_{B\neq A} \left[-K_{BA}P_{A}(t)+K_{AB}P_{B}(t) \right],
\end{equation}
where $K_{BA}$ is the rate constant of excitation transfer from $\ket{A}$ to $\ket{B}$ given by

\begin{equation}\label{eq:rate}
    \begin{aligned}
        K_{BA}&=\frac{2|V_{AB}|^{2}}{\hbar^{2}}\mathcal{R} \int_{0}^{\infty} \exp \left[-\frac{i t'}{\hbar} \left(E_{B}-E_{A}+\Lambda_{A}+\Lambda_{B} \right)  \right]\\
        &\times \exp \left[-g_{A}(t')-g_{B}(t') \right]dt'.
    \end{aligned}
\end{equation}
Here, $E_{A}$ and $E_{B}$ represent the electronic energy gaps from the ground to the excited state at chromophores $\ket{A}$ and $\ket{B}$, respectively, and $V_{AB}$ the electronic coupling between these chromophores. The total electron-vibrational coupling at a given chromophore $\ket{A}$ is measured by the reorganization energy $\Lambda_{A}$. The reorganization energy is related to the environment spectral density by
\begin{equation}
\Lambda_{A}=\int_{0}^{\infty}\frac{J_{A}(\omega)}{\omega}\: d\omega.
\end{equation}
The line-broadening function of chromophore $A$ is determined by
\begin{equation}
\begin{split}
        g_{A}(t)= \frac{1}{\hbar}\int_{0}^{\infty}J_{A}(\omega)\bigg[&\coth \left(\frac{\beta\hbar\omega}{2}\right)\frac{1-\cos(\omega t)}{\omega^{2}}\\+
        &i \frac{\sin(\omega t)-\omega t}{\omega^{2}} \bigg]
\end{split}
\end{equation}
where $\beta = 1/k_{B}T$ is the inverse thermal energy. 

Within this framework, the dissipation rate $\dot{E}_{Aj}(t)$ in the $j$-th vibrational mode belonging to the chromophore $\ket{A}$ can be evaluated as

\begin{equation}
    \dot{E}_{Aj}(t)= \sum_{B \neq A} \left[\mathcal{K}_{BA}^{Aj}P_{A}(t)+\mathcal{K}_{AB}^{Aj}P_{B}(t) \right]
\end{equation}
where the dissipation rate constants $\{\mathcal{K}_{BA}^{Aj}\}$ are computed as

\begin{equation}\label{eq:D-Rate}
        \mathcal{K}_{BA}^{Aj}=\frac{2|V_{AB}|^{2}}{\hbar^{2}} \lambda_{A}^{j} \mathcal{I}_{BA}(\omega_{Aj}).
\end{equation}
Here, $\lambda_{A}^{j}$ is the discrete reorganization energy of the $j$-th vibration that belongs to $\ket{A}$. If we sum over all the vibrations at chromophore $\ket{A}$ we obtain the total reorganization energy $\Lambda_{A}=\sum_{j}\lambda_{A}^{j}$. In turn, $\mathcal{I}_{BA}(\omega)$ is the ``dissipative potential'' defined as
\begin{equation}\label{eq:Diss-I}
\begin{aligned}
    \mathcal{I}_{BA}(\omega)=&\mathcal{R} \int_{0}^{\infty} \exp \left[-\frac{i t'}{\hbar} \left(E_{B}-E_{A}+\Lambda_{A}+\Lambda_{B} \right)  \right]\\
        &\times \exp \left[-g_{A}(t')-g_{B}(t') \right]\\
        &\times \left[\cos (\omega t')-i\coth\left(\frac{\beta\hbar\omega}{2}\right)\sin(\omega t')\right]dt'
\end{aligned}
\end{equation}
which quantifies the ability of the $j$-th vibration to induce dissipation per unit of reorganization energy at chromophore $\ket{A}$. To obtain an expression for the rate of dissipation at chromophore $\ket{A}$ as a function of frequency at a certain time, $\mathcal{D}_{A}(\omega,t)$, we replace $\mathcal{K}_{BA}^{Aj}$ and $\mathcal{K}_{AB}^{Aj}$ by the ``dissipative spectral densities'' $\mathcal{J}_{BA}^{A}(\omega)$ and $\mathcal{J}_{AB}^{A}(\omega)$
\begin{equation}\label{eq:Diss-J}
\begin{split}
    \mathcal{J}_{BA}^{A}(\omega)&=\frac{2|V_{AB}|^{2}}{\hbar^{2}} \frac{J_{A}(\omega)}{\omega} \mathcal{I}_{BA} (\omega)\\
    \mathcal{J}_{AB}^{A}(\omega)&=\frac{2|V_{AB}|^{2}}{\hbar^{2}} \frac{J_{A}(\omega)}{\omega} \mathcal{I}_{AB} (\omega)
\end{split}
\end{equation}
where we have used the property $\sum_{j}\lambda_{A}^{j}=\int_{0}^{\infty}d\omega\:J_{A}(\omega)/\omega $. This replacement yields the dissipation function
\begin{equation}\label{eq:Diss-rate}
    \mathcal{D}_{A}(\omega,t)=\sum_{B\neq A}\left[\mathcal{J}_{BA}^{A}(\omega)P_{A}(t)+\mathcal{J}_{AB}^{A}(\omega)P_{B}(t) \right].
\end{equation}
This equation enables us to compute the dissipation of chromophore $\ket{A}$ as a function of time and environment frequency. Importantly, all the terms in \eq{eq:Diss-rate} can be obtained without explicitly accessing the dynamics of the vibrational environment, keeping the computational cost tractable.   

The accumulated chromophore dissipation at a given time, $\mathcal{E}_{A}(\omega,t)$, can then be obtained as 
\begin{equation}\label{eq:Site-D}
    \mathcal{E}_{A}(\omega,t) = \int_{0}^{t} \mathcal{D}_{A}(\omega,t') dt'.
\end{equation}
In turn, the total time-dependent dissipation can be obtained as
\begin{equation}\label{eq:Total-D}
    \mathcal{E}_{\text{Tot}}(\omega,t)=\sum\limits_{A=1}^{N}\mathcal{E}_{A}(\omega,t),
\end{equation}
where $N$ is the number of chromophores.

\textbf{Non-Markovian Effects.} \Eq{eq:population} is a Markovian quantum master equation which assumes that the dynamics of the environment is fast compared to the system's dynamics. In our simulations, we account for non-Markovian effects that go beyond this treatment by using a time scale separation method\cite{berkelbach2012reduced,montoya2015extending} to split the spectral density, $J(\omega)$, into ``slow" and ``fast" components, denoted as $J_{\text{slow}}(\omega)$ and $J_{\text{fast}}(\omega)$, respectively. To introduce the desired non-Markovianity, we prevent the environment modes in $J_{\text{slow}}(\omega)$ from directly influencing the system dynamics. The spectral density separation is formally achieved by defining
\begin{equation}
    \begin{split}
        J_{\text{slow}}(\omega)&= S(\omega,\omega^{*})J(\omega)\\
        J_{\text{fast}}(\omega)&=[1-S(\omega,\omega^{*})]J(\omega)
    \end{split}
\end{equation}
where $S(\omega,\omega^{*})$ is the splitting function given by
\begin{equation}
    S(\omega,\omega^{*})=\left\{ \begin{array}{ccc} [1-(\omega/\omega^{*})^{2}]^{2}, &  & \omega<\omega^{*}\\  0, &  & \omega\geq\omega^{*} \end{array}\right.
\end{equation}
where $\omega^{*}$ is the cutoff frequency.

Next, we treat $J_{\text{slow}}(\omega)$ as a source of quasi-static noise that modulates the Bchl's electronic energies through a Gaussian random noise process with a standard deviation
\begin{equation}\label{eq:static}
    \sigma_{\text{slow}}=\frac{1}{\hbar} \int_{0}^{\infty} J_{\text{slow}}(\omega)\coth\left(\frac{\beta \hbar\omega}{2}\right)d\omega
\end{equation}
while $J_{\text{fast}}(\omega)$ is explicitly included in the dissipation dynamics in each noise realization as our spectral density. The final result is calculated by averaging the results over many noise realizations.

\textbf{Numerical Implementation.} The computation of dissipation pathways within the FMO complex requires the partitioning of each chromophore's spectral density into slow ($J_{\text{slow}}(\omega)$) and fast ($J_{\text{fast}}(\omega)$) components. A uniform cutoff frequency, $\omega^{*} = 20$ cm$^{-1}$, was applied across all bacteriochlorophylls (Bchls). This specific value was chosen to ensure that less than 5$\%$ of the total reorganization energy for each Bchl was attributed to slow-bath component, allowing the majority of the electron-vibrational coupling to be treated explicitly within the simulations. Benchmark HEOM computations (Fig. S3 in the SI) show that  this choice adequately captures population transfer rates and steady-state populations even for long-times $>$ 1 ps when the dynamics of some of these slow components of the bath can potentially play a role.

To account for the non-Markovian effects, we performed an ensemble average of over 10,000 independent noise realizations. Importantly, while each realization featured distinct site energies ($E$) drawn from a distribution representing slow-bath component, \eq{eq:static} , the underlying set of fast spectral density components ($J_{\text{fast}}(\omega)$) remained the same across all realizations. This significantly reduces the computational burden, as several crucial quantities, including the line-broadening functions ($g(t)$), total reorganization energies ($\Lambda$), and the discretized reorganization energies ($\lambda^{j}$ for each $j$-th vibrational mode), needed to be computed only once. Furthermore, to enhance computational efficiency, the calculation of the dissipative potential Eq. \ref{eq:Diss-I} was vectorized with respect to both the frequency ($\omega$) and time ($t$) grids using the NumPy library in Python. This vectorization allows for the simultaneous evaluation of these quantities for all frequency modes and time points, thereby significantly reducing the overall computation time.

The fast spectral density components, $J_{\text{fast}}(\omega)$, were discretized using 4000 effective harmonic oscillator modes equally spaced in frequency. The time integrals required for calculating the population transfer rate constants (Eq. \ref{eq:rate}) and the dissipation rates (Eq. \ref{eq:Diss-J}) were evaluated numerically using the trapezoidal rule. This method was applied over a finely spaced time grid, with a step size of $\Delta t = 0.5 \, \text{fs}$ and extending to a maximum integration time of $T_{\text{max}} = 30 \, \text{ps}$. The trapezoidal rule provides a good balance between accuracy and computational cost for these types of integrals. The propagation of the rate equations governing the electronic populations (Eq. \ref{eq:population}) and the dissipation dynamics (Eq. \ref{eq:D-Rate}) was performed using the fourth-order Runge-Kutta (RK4) method.  A consistent time step of $0.5 \, \text{fs}$ was employed for the RK4 integration, ensuring numerical stability and accuracy throughout the simulation. The accuracy of this approach in generating semi-quantitative results for the FMO complex is assessed against the numerically exact HEOM using illustrative models (see Secs. SII-SIII of the Supplementary Information). The codes used to obtain the results of this paper are available on GitHub.\cite{GustinGitHub}
\break
\break
\break
\break
\break

CWK was supported by the National Research Foundation of Korea (NRF) grant funded by the Ministry of Science and ICT (MSIT) of Korea (Grant Number: 2022R1F1A1074027, 2023M3K5A1094813, and RS-2023-00218219). I.G and I.F are supported by the U.S. Department of Energy, Office of Science, Office of Basic Energy Sciences, Quantum Information Science Research in Chemical Sciences, Geosciences, and Biosciences Program under Award Number DE-SC0025334.

The Supporting Information provides details regarding the applicability of the dissipation pathways theory to the Fenna-Matthews-Olson (FMO) complex.  Additional information in the Supporting Information includes the electronic Hamiltonian for the FMO complex, the spectral densities used, the methodology for extracting the vibrational modes, and the dissipation dynamics for a different initial condition. We also present a set of animations for the extracted vibrational modes around 200 cm$^{-1}$ in each bacteriochlorophyll.

\bibliography{achemso-demo}

@book{blankenship2021molecular,
  title={Molecular mechanisms of photosynthesis},
  author={Blankenship, Robert E},
  year={2021},
  publisher={John Wiley \& Sons}}

@article{baker2008chlorophyll,
  title={Chlorophyll fluorescence: a probe of photosynthesis in vivo},
  author={Baker, Neil R},
  journal={Annu. Rev. Plant Biol.},
  volume={59},
  pages={89--113},
  year={2008},
  publisher={Annual Reviews}}

@article{mirkovic2017light,
  title={Light absorption and energy transfer in the antenna complexes of photosynthetic organisms},
  author={Mirkovic, Tihana and Ostroumov, Evgeny E and Anna, Jessica M and Van Grondelle, Rienk and Govindjee and Scholes, Gregory D},
  journal={Chem. Rev.},
  volume={117},
  number={2},
  pages={249--293},
  year={2017}}

@article{allodi2018redox,
  title={Redox conditions affect ultrafast exciton transport in photosynthetic pigment--protein complexes},
  author={Allodi, Marco A and Otto, John P and Sohail, Sara H and Saer, Rafael G and Wood, Ryan E and Rolczynski, Brian S and Massey, Sara C and Ting, Po-Chieh and Blankenship, Robert E and Engel, Gregory S},
  journal={J. Phys. Chem. Lett.},
  volume={9},
  number={1},
  pages={89--95},
  year={2018},
  publisher={ACS Publications}
}

@article{brixner2005two,
  title={Two-dimensional spectroscopy of electronic couplings in photosynthesis},
  author={Brixner, Tobias and Stenger, Jens and Vaswani, Harsha M and Cho, Minhaeng and Blankenship, Robert E and Fleming, Graham R},
  journal={Nature},
  volume={434},
  number={7033},
  pages={625--628},
  year={2005},
  publisher={Nature Publishing Group UK London}
}

@article{engel2007evidence,
  title={Evidence for wavelike energy transfer through quantum coherence in photosynthetic systems},
  author={Engel, Gregory S and Calhoun, Tessa R and Read, Elizabeth L and Ahn, Tae-Kyu and Man{\v{c}}al, Tom{\'a}{\v{s}} and Cheng, Yuan-Chung and Blankenship, Robert E and Fleming, Graham R},
  journal={Nature},
  volume={446},
  number={7137},
  pages={782--786},
  year={2007},
  publisher={Nature Publishing Group UK London}
}

@article{panitchayangkoon2010long,
  title={Long-lived quantum coherence in photosynthetic complexes at physiological temperature},
  author={Panitchayangkoon, Gitt and Hayes, Dugan and Fransted, Kelly A and Caram, Justin R and Harel, Elad and Wen, Jianzhong and Blankenship, Robert E and Engel, Gregory S},
  journal={ Proc. Natl. Acad. Sci.},
  volume={107},
  number={29},
  pages={12766--12770},
  year={2010},
  publisher={National Acad Sciences}
}

@article{thyrhaug2016exciton,
  title={Exciton structure and energy transfer in the Fenna--Matthews--Olson complex},
  author={Thyrhaug, Erling and Zidek, Karel and Dost{\'a}l, Jakub and B{\'\i}na, David and Zigmantas, Donatas},
  journal={J. Phys. Chem. Lett.},
  volume={7},
  number={9},
  pages={1653--1660},
  year={2016},
  publisher={ACS Publications}
}

@article{dostal2016situ,
  title={In situ mapping of the energy flow through the entire photosynthetic apparatus},
  author={Dost{\'a}l, Jakub and P{\v{s}}en{\v{c}}{\'\i}k, Jakub and Zigmantas, Donatas},
  journal={Nat. Chem.},
  volume={8},
  number={7},
  pages={705--710},
  year={2016},
  publisher={Nature Publishing Group UK London}
}

@article{duan2017nature,
  title={Nature does not rely on long-lived electronic quantum coherence for photosynthetic energy transfer},
  author={Duan, Hong-Guang and Prokhorenko, Valentyn I and Cogdell, Richard J and Ashraf, Khuram and Stevens, Amy L and Thorwart, Michael and Miller, RJ Dwayne},
  journal={Proc. Natl. Acad. Sci.},
  volume={114},
  number={32},
  pages={8493--8498},
  year={2017},
  publisher={National Acad Sciences}
}

@article{maiuri2018coherent,
  title={Coherent wavepackets in the Fenna--Matthews--Olson complex are robust to excitonic-structure perturbations caused by mutagenesis},
  author={Maiuri, Margherita and Ostroumov, Evgeny E and Saer, Rafael G and Blankenship, Robert E and Scholes, Gregory D},
  journal={Nat. chem.},
  volume={10},
  number={2},
  pages={177--183},
  year={2018},
  publisher={Nature Publishing Group UK London}
}

@article{thyrhaug2018identification,
  title={Identification and characterization of diverse coherences in the Fenna--Matthews--Olson complex},
  author={Thyrhaug, Erling and Tempelaar, Roel and Alcocer, Marcelo JP and {\v{Z}}{\'\i}dek, Karel and B{\'\i}na, David and Knoester, Jasper and Jansen, Thomas LC and Zigmantas, Donatas},
  journal={Nat. chem.},
  volume={10},
  number={7},
  pages={780--786},
  year={2018},
  publisher={Nature Publishing Group}
}

@article{kim2018excited,
  title={Excited state energy fluctuations in the Fenna--Matthews--Olson complex from molecular dynamics simulations with interpolated chromophore potentials},
  author={Kim, Chang Woo and Choi, Bongsik and Rhee, Young Min},
  journal={Phys. Chem. Chem. Phys.},
  volume={20},
  number={5},
  pages={3310--3319},
  year={2018},
  publisher={Royal Society of Chemistry}
}

@article{moix2011efficient,
  title={Efficient energy transfer in light-harvesting systems, III: The influence of the eighth bacteriochlorophyll on the dynamics and efficiency in FMO},
  author={Moix, Jeremy and Wu, Jianlan and Huo, Pengfei and Coker, David and Cao, Jianshu},
  journal={J. Phys. Chem. Lett.},
  volume={2},
  number={24},
  pages={3045--3052},
  year={2011},
  publisher={ACS Publications}
}

@article{blau2018local,
  title={Local protein solvation drives direct down-conversion in phycobiliprotein PC645 via incoherent vibronic transport},
  author={Blau, Samuel M and Bennett, Doran IG and Kreisbeck, Christoph and Scholes, Gregory D and Aspuru-Guzik, Al{\'a}n},
  journal={Proc. Natl. Acad. Sci.},
  volume={115},
  number={15},
  pages={E3342--E3350},
  year={2018},
  publisher={National Acad Sciences}
}

@article{hart2021engineering,
  title={Engineering couplings for exciton transport using synthetic DNA scaffolds},
  author={Hart, Stephanie M and Chen, Wei Jia and Banal, James L and Bricker, William P and Dodin, Amro and Markova, Larysa and Vyborna, Yuliia and Willard, Adam P and H{\"a}ner, Robert and Bathe, Mark and others},
  journal={Chem},
  volume={7},
  number={3},
  pages={752--773},
  year={2021},
  publisher={Elsevier}
}

@article{tang2024simulating,
  title={Simulating photosynthetic energy transport on a photonic network},
  author={Tang, Hao and Shang, Xiao-Wen and Shi, Zi-Yu and He, Tian-Shen and Feng, Zhen and Wang, Tian-Yu and Shi, Ruoxi and Wang, Hui-Ming and Tan, Xi and Xu, Xiao-Yun and others},
  journal={Npj Quantum Inf.},
  volume={10},
  number={1},
  pages={29},
  year={2024},
  publisher={Nature Publishing Group UK London}
}

@article{wasielewski2009self,
  title={Self-assembly strategies for integrating light harvesting and charge separation in artificial photosynthetic systems},
  author={Wasielewski, Michael R},
  journal={Acc. Chem. Res.},
  volume={42},
  number={12},
  pages={1910--1921},
  year={2009},
  publisher={ACS Publications}
}

@article{Kienzler2014,
	author = {Kienzler, D. and Lo, H.-Y. and Keitch, B. and de Clercq, L. and Leupold, F. and Lindenfelser, F. and Marinelli, M. and Negnevitsky, V and Home, J. P.},
	doi = {10.1126/science.1261033},
	journal = {Science},
	number = {6217},
	pages = {53-56},
	title = {Quantum Harmonic Oscillator State Synthesis by Reservoir Engineering},
	volume = {347},
	year = {2014}}

@article{Hsu2017,
	author = {Hsu, Liang-Yan and Ding, Wendu and Schatz, George C.},
	doi = {10.1021/acs.jpclett.7b00526},
	journal = {J. Phys. Chem. Lett.},
	number = {10},
	pages = {2357-2367},
	title = {Plasmon-Coupled Resonance Energy Transfer},
	volume = {8},
	year = {2017}}

@article{Ng2020,
	author = {Ng, Kara and Webster, Megan and Carbery, William P. and Visaveliya, Nikunjkumar and Gaikwad, Pooja and Jang, Seogjoo J. and Kretzschmar, Ilona and Eisele, Dorthe M.},
	doi = {10.1038/s41557-020-00563-4},
	journal = {Nat. Chem.},
	number = {12},
	pages = {1157-1164},
	title = {Frenkel Excitons in Heat-Stressed Supramolecular Nanocomposites Enabled by Tunable Cage-Like Scaffolding},
	volume = {12},
	year = {2020}}

@article{CamposGonzalezAngulo2019,
	author = {Campos-Gonzalez-Angulo, Jorge A. and Ribeiro, Raphael F. and Yuen-Zhou, Joel},
	doi = {10.1038/s41467-019-12636-1},
	journal = {Nat. Commun.},
	number = {1},
	pages = {4685},
	title = {Resonant Catalysis of Thermally Activated Chemical Reactions with Vibrational Polaritons},
	volume = {10},
	year = {2019}}

@article{gustin2023mapping,
  title={Mapping electronic decoherence pathways in molecules},
  author={Gustin, Ignacio and Kim, Chang Woo and McCamant, David W and Franco, Ignacio},
  journal={Proc. Natl. Acad. Sci.},
  volume={120},
  number={49},
  pages={e2309987120},
  year={2023},
  publisher={National Acad Sciences}
}

@article{ratsep2007electron,
  title={Electron--phonon and vibronic couplings in the FMO bacteriochlorophyll a antenna complex studied by difference fluorescence line narrowing},
  author={R{\"a}tsep, Margus and Freiberg, Arvi},
  journal={J. Lumin.},
  volume={127},
  number={1},
  pages={251--259},
  year={2007},
  publisher={Elsevier}
}

@article{novoderezhkin2004energy,
  title={Energy-transfer dynamics in the LHCII complex of higher plants: modified redfield approach},
  author={Novoderezhkin, Vladimir I and Palacios, Miguel A and Van Amerongen, Herbert and Van Grondelle, Rienk},
  journal={J. Phys. Chem. B},
  volume={108},
  number={29},
  pages={10363--10375},
  year={2004},
  publisher={ACS Publications}
}

@article{lee2017first,
  title={First-principles models for biological light-harvesting: phycobiliprotein complexes from cryptophyte algae},
  author={Lee, Mi Kyung and Bravaya, Ksenia B and Coker, David F},
  journal={J. Am. Chem. Soc.},
  volume={139},
  number={23},
  pages={7803--7814},
  year={2017},
  publisher={ACS Publications}
}

@article{fenna1975chlorophyll,
  title={Chlorophyll arrangement in a bacteriochlorophyll protein from Chlorobium limicola},
  author={Fenna, RE and Matthews, BW},
  journal={Nature},
  volume={258},
  number={5536},
  pages={573--577},
  year={1975},
  publisher={Nature Publishing Group UK London}
}

@article{adolphs2006proteins,
  title={How proteins trigger excitation energy transfer in the FMO complex of green sulfur bacteria},
  author={Adolphs, Julia and Renger, Thomas},
  journal={Biophys. J.},
  volume={91},
  number={8},
  pages={2778--2797},
  year={2006},
  publisher={Elsevier}
}

@article{higgins2021photosynthesis,
  title={Photosynthesis tunes quantum-mechanical mixing of electronic and vibrational states to steer exciton energy transfer},
  author={Higgins, Jacob S and Lloyd, Lawson T and Sohail, Sara H and Allodi, Marco A and Otto, John P and Saer, Rafael G and Wood, Ryan E and Massey, Sara C and Ting, Po-Chieh and Blankenship, Robert E and others},
  journal={Proc. Natl. Acad. Sci.},
  volume={118},
  number={11},
  pages={e2018240118},
  year={2021},
  publisher={National Acad Sciences}
}

@article{higgins2021redox,
  title={Redox conditions correlated with vibronic coupling modulate quantum beats in photosynthetic pigment--protein complexes},
  author={Higgins, Jacob S and Allodi, Marco A and Lloyd, Lawson T and Otto, John P and Sohail, Sara H and Saer, Rafael G and Wood, Ryan E and Massey, Sara C and Ting, Po-Chieh and Blankenship, Robert E and others},
  journal={Proc. Natl. Acad. Sci.},
  volume={118},
  number={49},
  pages={e2112817118},
  year={2021},
  publisher={National Acad Sciences}
}

@article{womick2011vibronic,
  title={Vibronic enhancement of exciton sizes and energy transport in photosynthetic complexes},
  author={Womick, Jordan M and Moran, Andrew M},
  journal={J. Phys. Chem. B},
  volume={115},
  number={6},
  pages={1347--1356},
  year={2011},
  publisher={ACS Publications}
}

@article{montoya2015extending,
  title={Extending the applicability of Redfield theories into highly non-Markovian regimes},
  author={Montoya-Castillo, Andr{\'e}s and Berkelbach, Timothy C and Reichman, David R},
  journal={J. Chem. Phys.},
  volume={143},
  number={19},
pages = {194108},
  year={2015},
  publisher={AIP Publishing}
}

@article{berkelbach2012reduced,
  title={Reduced density matrix hybrid approach: Application to electronic energy transfer},
  author={Berkelbach, Timothy C and Markland, Thomas E and Reichman, David R},
  journal={J. Chem. Phys.},
  volume={136},
pages={084104},
  number={8},
  year={2012},
  publisher={AIP Publishing}
}

@article{jang2018delocalized,
  title={Delocalized excitons in natural light-harvesting complexes},
  author={Jang, Seogjoo J and Mennucci, Benedetta},
  journal={Rev. Mod. Phys.},
  volume={90},
  number={3},
  pages={035003},
  year={2018},
  publisher={APS}
}

@article{cao2020quantum,
  title={Quantum biology revisited},
  author={Cao, Jianshu and Cogdell, Richard J and Coker, David F and Duan, Hong-Guang and Hauer, J{\"u}rgen and Kleinekath{\"o}fer, Ulrich and Jansen, Thomas LC and Man{\v{c}}al, Tom{\'a}{\v{s}} and Miller, RJ Dwayne and Ogilvie, Jennifer P and others},
  journal={Sci. Adv.},
  volume={6},
  number={14},
  pages={eaaz4888},
  year={2020},
  publisher={American Association for the Advancement of Science}
}

@article{tempo,
	author = {Strathearn, Aidan and Kirton, Peter and Kilda, Dainius and Keeling, Jonathan and Lovett, Brendon William},
	journal = {Nat. Commun.},
	number = {1},
	pages = {1--9},
	publisher = {Nature Publishing Group},
	title = {Efficient non-Markovian quantum dynamics using time-evolving matrix product operators},
	volume = {9},
	year = {2018}}

@article{Bose2022multisite,
	author = {Amartya Bose and Peter L. Walters},
	journal = {J. Chem. Phys.},
	pages = {024101},
	title = {A multisite decomposition of the tensor network path integrals},
	volume = {156},
	year = {2022}}

@article{arsenault2020role,
  title={The role of mixed vibronic Qy-Qx states in green light absorption of light-harvesting complex II},
  author={Arsenault, Eric A and Yoneda, Yusuke and Iwai, Masakazu and Niyogi, Krishna K and Fleming, Graham R},
  journal={Nat. Commun.},
  volume={11},
  number={1},
  pages={6011},
  year={2020},
  publisher={Nature Publishing Group UK London}
}

@article{lambrev2020insights,
  title={Insights into the mechanisms and dynamics of energy transfer in plant light-harvesting complexes from two-dimensional electronic spectroscopy},
  author={Lambrev, Petar H and Akhtar, Parveen and Tan, Howe-Siang},
  journal={Biochim. Biophys. Acta Bioenerg.},
  volume={1861},
  number={4},
  pages={148050},
  year={2020},
  publisher={Elsevier}
}

@article{biswas2022coherent,
  title={Coherent two-dimensional and broadband electronic spectroscopies},
  author={Biswas, Somnath and Kim, JunWoo and Zhang, Xinzi and Scholes, Gregory D},
  journal={Chem. Rev.},
  volume={122},
  number={3},
  pages={4257--4321},
  year={2022},
  publisher={ACS Publications}
}

@article{wiethorn2023beyond,
  title={Beyond the Condon limit: Condensed phase optical spectra from atomistic simulations},
  author={Wiethorn, Zachary R and Hunter, Kye E and Zuehlsdorff, Tim J and Montoya-Castillo, Andr{\'e}s},
  journal={J. Chem. Phys.},
  volume={159},
  number={24},
pages = {244114},
  year={2023},
  publisher={AIP Publishing}
}

@article{hu2022tuning,
  title={Tuning and enhancing quantum coherence time scales in molecules via light-matter hybridization},
  author={Hu, Wenxiang and Gustin, Ignacio and Krauss, Todd D and Franco, Ignacio},
  journal={J. Phys. Chem. Lett.},
  volume={13},
  number={49},
  pages={11503--11511},
  year={2022},
  publisher={ACS Publications}
}

@article{sagnella2000time,
  title={Time scales and pathways for kinetic energy relaxation in solvated proteins: Application to carbonmonoxy myoglobin},
  author={Sagnella, Diane E and Straub, John E and Thirumalai, D},
  journal={J. Chem. Phys.},
  volume={113},
  number={17},
  pages={7702--7711},
  year={2000},
  publisher={AIP Publishing}
}

@article{bu2003simulating,
  title={Simulating vibrational energy flow in proteins: relaxation rate and mechanism for heme cooling in cytochrome c},
  author={Bu, Lintao and Straub, John E},
  journal={J. Phys. Chem. B},
  volume={107},
  number={44},
  pages={12339--12345},
  year={2003},
  publisher={ACS Publications}
}

@article{ishikura2015energy,
  title={Energy exchange network of inter-residue interactions within a thermally fluctuating protein molecule: a computational study},
  author={Ishikura, Takakazu and Iwata, Yuki and Hatano, Tatsuro and Yamato, Takahisa},
  journal={J. Comput. Chem.},
  volume={36},
  number={22},
  pages={1709--1718},
  year={2015},
  publisher={Wiley Online Library}
}

@article{xu2014vibrational,
  title={Vibrational energy flow through the green fluorescent protein--water interface: communication maps and thermal boundary conductance},
  author={Xu, Yao and Leitner, David M},
  journal={J. Phys. Chem. B},
  volume={118},
  number={28},
  pages={7818--7826},
  year={2014},
  publisher={ACS Publications}
}

@article{xu2014communication,
  title={Communication maps of vibrational energy transport through Photoactive Yellow Protein},
  author={Xu, Yao and Leitner, David M},
  journal={J. Phys. Chem. A},
  volume={118},
  number={35},
  pages={7280--7287},
  year={2014},
  publisher={ACS Publications}
}

@article{leitner2015vibrational,
  title={Vibrational energy flow in the villin headpiece subdomain: Master equation simulations},
  author={Leitner, David M and Buchenberg, Sebastian and Brettel, Paul and Stock, Gerhard},
  journal={J. Chem. Phys.},
  volume={142},
  pages={075101},
  number={7},
  year={2015},
  publisher={AIP Publishing}
}

@article{leitner2009frequency,
  title={Frequency-resolved communication maps for proteins and other nanoscale materials},
  author={Leitner, David M},
  journal={J. Chem. Phys.},
  volume={130},
  number={19},
  pages={195101},
  year={2009},
  publisher={AIP Publishing}
}

@article{agbo2015vibrational,
  title={Vibrational energy flow across heme--cytochrome c and cytochrome c--water interfaces},
  author={Agbo, Johnson K and Xu, Yao and Zhang, Ping and Straub, John E and Leitner, David M},
  journal={Theor. Chem. Acc.},
volume = {133},
  pages={129--138},
  year={2015},
  publisher={Springer}
}

@article{segatta_quantum_2017,
    title = {A {Quantum} {Chemical} {Interpretation} of {Two}-{Dimensional} {Electronic} {Spectroscopy} of {Light}-{Harvesting} {Complexes}},
    volume = {139},
    issn = {0002-7863},
    url = {https://doi.org/10.1021/jacs.7b02130},
    doi = {10.1021/jacs.7b02130},
    number = {22},
    urldate = {2024-10-22},
    journal = {J. Am. Chem. Soc.},
    author = {Segatta, Francesco and Cupellini, Lorenzo and Jurinovich, Sandro and Mukamel, Shaul and Dapor, Maurizio and Taioli, Simone and Garavelli, Marco and Mennucci, Benedetta},
    month = jun,
    year = {2017},
    
    pages = {7558--7567},
}

@article{pereverzev2009energy,
  title={Energy and charge-transfer dynamics using projected modes},
  author={Pereverzev, Andrey and Bittner, Eric R and Burghardt, Irene},
  journal={J. Chem. Phys.},
  volume={131},
  number={3},
  year={2009},
  pages = {034104}
}

@article{yang2017identifying,
  title={Identifying electron transfer coordinates in donor-bridge-acceptor systems using mode projection analysis},
  author={Yang, Xunmo and Keane, Theo and Delor, Milan and Meijer, Anthony JHM and Weinstein, Julia and Bittner, Eric R},
  journal={Nat. Commun.},
  volume={8},
  number={1},
  pages={14554},
  year={2017},
  publisher={Nature Publishing Group UK London}
}

@article{abramavicius_excitation_2014,
    title = {Excitation transfer pathways in excitonic aggregates revealed by the stochastic {Schrödinger} equation},
    volume = {140},
    issn = {0021-9606},
    url = {https://doi.org/10.1063/1.4863968},
    doi = {10.1063/1.4863968},
    number = {6},
    urldate = {2025-03-25},
    journal = {J. Chem. Phys.},
    author = {Abramavicius, Vytautas and Abramavicius, Darius},
    month = feb,
    year = {2014},
    pages = {065103},
}

@article{maly_role_2016,
    title = {The {Role} of {Resonant} {Vibrations} in {Electronic} {Energy} {Transfer}},
    volume = {17},
    copyright = {© 2016 The Authors. Published by Wiley-VCH Verlag GmbH \& Co. KGaA.},
    issn = {1439-7641},
    url = {https://onlinelibrary.wiley.com/doi/abs/10.1002/cphc.201500965},
    doi = {10.1002/cphc.201500965},
    language = {en},
    number = {9},
    urldate = {2025-03-25},
    journal = {Chem. Phys. Chem.},
    author = {Malý, Pavel and Somsen, Oscar J. G. and Novoderezhkin, Vladimir I. and Mančal, Tomáš and van Grondelle, Rienk},
    year = {2016},
    pages = {1356--1368},
}

@article{olbrich_theory_2011,
    title = {Theory and {Simulation} of the {Environmental} {Effects} on {FMO} {Electronic} {Transitions}},
    volume = {2},
    url = {https://doi.org/10.1021/jz2007676},
    doi = {10.1021/jz2007676},
    number = {14},
    urldate = {2025-03-25},
    journal = {J. Phys. Chem. Lett.},
    author = {Olbrich, Carsten and Strümpfer, Johan and Schulten, Klaus and Kleinekath{\"o}fer, Ulrich},
    month = jul,
    year = {2011},
    
    pages = {1771--1776},
}

@article{schulze_explicit_2015,
    title = {Explicit {Correlated} {Exciton}-{Vibrational} {Dynamics} of the {FMO} {Complex}},
    volume = {119},
    issn = {1520-6106},
    url = {https://doi.org/10.1021/acs.jpcb.5b03928},
    doi = {10.1021/acs.jpcb.5b03928},
    number = {20},
    urldate = {2025-03-25},
    journal = {J. Phys. Chem. B},
    author = {Schulze, J. and Kühn, O.},
    month = may,
    year = {2015},
    
    pages = {6211--6216},
}

@article{padula_chromophore-dependent_2017,
    title = {Chromophore-{Dependent} {Intramolecular} {Exciton}–{Vibrational} {Coupling} in the {FMO} {Complex}: {Quantification} and {Importance} for {Exciton} {Dynamics}},
    volume = {121},
    issn = {1520-6106},
    shorttitle = {Chromophore-{Dependent} {Intramolecular} {Exciton}–{Vibrational} {Coupling} in the {FMO} {Complex}},
    url = {https://doi.org/10.1021/acs.jpcb.7b08020},
    doi = {10.1021/acs.jpcb.7b08020},
    number = {43},
    urldate = {2025-03-25},
    journal = {J. Phys. Chem. B},
    author = {Padula, Daniele and Lee, Myeong H. and Claridge, Kirsten and Troisi, Alessandro},
    month = nov,
    year = {2017},
    
    pages = {10026--10035},
}

@article{wu_efficient_2012,
    title = {Efficient energy transfer in light-harvesting systems: {Quantum}-classical comparison, flux network, and robustness analysis},
    volume = {137},
    issn = {0021-9606},
    shorttitle = {Efficient energy transfer in light-harvesting systems},
    url = {https://doi.org/10.1063/1.4762839},
    doi = {10.1063/1.4762839},
    number = {17},
    urldate = {2025-03-25},
    journal = {J. Chem. Phys.},
    author = {Wu, Jianlan and Liu, Fan and Ma, Jian and Silbey, Robert J. and Cao, Jianshu},
    month = nov,
    year = {2012},
    pages = {174111},
}

@article{gustin_decoherence_2025,
    title = {Decoherence dynamics in molecular qubits: exponential, gaussian and beyond},
    volume = {162},
    issn = {0021-9606},
    shorttitle = {Decoherence dynamics in molecular qubits},
    url = {https://doi.org/10.1063/5.0246970},
    doi = {10.1063/5.0246970},
    language = {en},
    number = {6},
    urldate = {2025-03-25},
    journal = {J. Chem. Phys.},
    author = {Gustin, Ignacio and Chen, Xinxian and Franco, Ignacio},
    month = feb,
    year = {2025},
    pages = {64106},
}

@article{lee_modeling_2016,
	title = {Modeling {Electronic}-{Nuclear} {Interactions} for {Excitation} {Energy} {Transfer} {Processes} in {Light}-{Harvesting} {Complexes}},
	volume = {7},
	url = {https://doi.org/10.1021/acs.jpclett.6b01440},
	doi = {10.1021/acs.jpclett.6b01440},
	number = {16},
	urldate = {2022-02-02},
	journal = {J. Phys. Chem. Lett.},
	author = {Lee, Mi Kyung and Coker, David F.},
	month = aug,
	year = {2016},
	
	pages = {3171--3178},
}

@article{kim2021theory,
	title = {Theory of dissipation pathways in open quantum systems},
	volume = {154},
	issn = {0021-9606},
	url = {https://aip.scitation.org/doi/10.1063/5.0038967},
	doi = {10.1063/5.0038967},
	number = {8},
	urldate = {2023-03-14},
	journal = {J. Chem. Phys.},
	author = {Kim, Chang Woo and Franco, Ignacio},
	month = feb,
	year = {2021},
	
	pages = {084109},
}

@article{rivera_influence_2013,
	title = {Influence of {Site}-{Dependent} {Pigment}–{Protein} {Interactions} on {Excitation} {Energy} {Transfer} in {Photosynthetic} {Light} {Harvesting}},
	volume = {117},
	issn = {1520-6106},
	url = {https://doi.org/10.1021/jp4011586},
	doi = {10.1021/jp4011586},
	number = {18},
	urldate = {2023-07-13},
	journal = {J. Phys. Chem. B},
	author = {Rivera, Eva and Montemayor, Daniel and Masia, Marco and Coker, David F.},
	month = may,
	year = {2013},
	
	pages = {5510--5521},
}

@article{gillis_theoretical_2015,
	title = {A {Theoretical} {Investigation} into the {Effects} of {Temperature} on {Spatiotemporal} {Dynamics} of {EET} in the {FMO} {Complex}},
	volume = {119},
	issn = {1520-6106},
	url = {https://doi.org/10.1021/jp509103e},
	doi = {10.1021/jp509103e},
	number = {11},
	urldate = {2023-07-13},
	journal = {J. Phys. Chem. B},
	author = {Gillis, Colm G. and Jones, Garth A.},
	month = mar,
	year = {2015},
	
	pages = {4165--4174},
}

@article{christensson_origin_2012,
	title = {Origin of {Long}-{Lived} {Coherences} in {Light}-{Harvesting} {Complexes}},
	volume = {116},
	issn = {1520-6106},
	url = {https://doi.org/10.1021/jp304649c},
	doi = {10.1021/jp304649c},
	number = {25},
	urldate = {2023-07-18},
	journal = {J. Phys. Chem. B},
	author = {Christensson, Niklas and Kauffmann, Harald F. and Pullerits, Tõnu and Mančal, Tomáš},
	month = jun,
	year = {2012},
	
	pages = {7449--7454},
}

@article{wendling_electronvibrational_2000,
	title = {Electron-{Vibrational} {Coupling} in the {Fenna}-{Matthews}-{Olson} {Complex} of {Prosthecochloris} aestuarii {Determined} by {Temperature}-{Dependent} {Absorption} and {Fluorescence} {Line}-{Narrowing} {Measurements}},
	volume = {104},
	issn = {1520-6106},
	doi = {10.1021/jp000077+},
	number = {24},
	urldate = {2023-08-29},
	journal = {J. Phys. Chem. B},
	author = {Wendling, Markus and Pullerits, Tõnu and Przyjalgowski, Milosz A. and Vulto, Simone I. E. and Aartsma, Thijs J. and van Grondelle, Rienk and van Amerongen, Herbert},
	month = jun,
	year = {2000},
	pages = {5825--5831},
}

@article{kell_effect_2016,
	title = {Effect of {Spectral} {Density} {Shapes} on the {Excitonic} {Structure} and {Dynamics} of the {Fenna}–{Matthews}–{Olson} {Trimer} from {Chlorobaculum} tepidum},
	volume = {120},
	issn = {1089-5639},
	url = {https://doi.org/10.1021/acs.jpca.6b03107},
	doi = {10.1021/acs.jpca.6b03107},
	number = {31},
	urldate = {2023-10-03},
	journal = {J. Phys. Chem. A},
	author = {Kell, Adam and Blankenship, Robert E. and Jankowiak, Ryszard},
	month = aug,
	year = {2016},
	
	pages = {6146--6154},
}

@article{orf_evidence_2016,
	title = {Evidence for a cysteine-mediated mechanism of excitation energy regulation in a photosynthetic antenna complex},
	volume = {113},
	url = {https://www.pnas.org/doi/abs/10.1073/pnas.1603330113},
	doi = {10.1073/pnas.1603330113},
	number = {31},
	urldate = {2023-10-11},
	journal = {Proc. Natl. Acad. Sci.},
	author = {Orf, Gregory S. and Saer, Rafael G. and Niedzwiedzki, Dariusz M. and Zhang, Hao and McIntosh, Chelsea L. and Schultz, Jason W. and Mirica, Liviu M. and Blankenship, Robert E.},
	month = aug,
	year = {2016},
	pages = {E4486--E4493},
}

@article{gonzalez-soria_parametric_2020,
	title = {Parametric {Mapping} of {Quantum} {Regime} in {Fenna}–{Matthews}–{Olson} {Light}-{Harvesting} {Complexes}: {A} {Synthetic} {Review} of {Models}, {Methods} and {Approaches}},
	volume = {10},
	copyright = {http://creativecommons.org/licenses/by/3.0/},
	issn = {2076-3417},
	shorttitle = {Parametric {Mapping} of {Quantum} {Regime} in {Fenna}–{Matthews}–{Olson} {Light}-{Harvesting} {Complexes}},
	url = {https://www.mdpi.com/2076-3417/10/18/6474},
	doi = {10.3390/app10186474},
	language = {en},
	number = {18},
	urldate = {2023-11-14},
	journal = {Appl. Sci.},
	author = {González-Soria, Bruno and Delgado, Francisco and Anaya-Morales, Alan},
	month = jan,
	year = {2020},
	pages = {6474},
}

@article{kim2024general1,
	title = {General framework for quantifying dissipation pathways in open quantum systems. {I}. {Theoretical} formulation},
	volume = {160},
	issn = {0021-9606},
	url = {https://doi.org/10.1063/5.0202860},
	doi = {10.1063/5.0202860},
	number = {21},
	urldate = {2024-10-22},
	journal = {J. Chem. Phys.},
	author = {Kim, Chang Woo and Franco, Ignacio},
	month = jun,
	year = {2024},
	pages = {214111},
}

@article{kim2024general2,
	title = {General framework for quantifying dissipation pathways in open quantum systems. {II}. {Numerical} validation and the role of non-{Markovianity}},
	volume = {160},
	issn = {0021-9606},
	url = {https://doi.org/10.1063/5.0202862},
	doi = {10.1063/5.0202862},
	number = {21},
	urldate = {2024-10-22},
	journal = {J. Chem. Phys.},
	author = {Kim, Chang Woo and Franco, Ignacio},
	month = jun,
	year = {2024},
	pages = {214112},
}

@article{kim_extracting_2022,
	title = {Extracting bath information from open-quantum-system dynamics with the hierarchical equations-of-motion method},
	volume = {106},
	url = {https://link.aps.org/doi/10.1103/PhysRevA.106.042223},
	doi = {10.1103/PhysRevA.106.042223},
	number = {4},
	urldate = {2024-10-22},
	journal = {Phys. Rev. A},
	author = {Kim, Chang Woo},
	month = oct,
	year = {2022},
	pages = {042223},
}

@article{kundu_tight_2022,
	title = {Tight inner ring architecture and quantum motion of nuclei enable efficient energy transfer in bacterial light harvesting},
	volume = {8},
	url = {https://www.science.org/doi/full/10.1126/sciadv.add0023},
	doi = {10.1126/sciadv.add0023},
	number = {43},
	urldate = {2024-10-22},
	journal = {Sci. Adv.},
	author = {Kundu, Sohang and Dani, Reshmi and Makri, Nancy},
	month = oct,
	year = {2022},
	pages = {eadd0023}
}

@misc{GustinGitHub,
  author    = {Ignacio Gustin and Chang Woo Kim and Ignacio Franco},
  title     = {Dissipation Pathways in a Photosynthetic Complex},
  year      = {2025},
  publisher = {GitHub},
  note      = {GitHub repository: \url{https://github.com/ifgroup/FMO_Dissipation} (accessed Oct 25, 2025)},
}

@article{tanimura2020numerically,
	author = {Tanimura, Yoshitaka},
	journal = {J. Chem. Phys.},
	number = {2},
	pages = {020901},
	publisher = {AIP Publishing LLC},
	title = {Numerically ``exact'' approach to open quantum dynamics: The hierarchical equations of motion (HEOM)},
	volume = {153},
	year = {2020}}

@article{Ikeda2020,
	author = {Ikeda,Tatsushi and Scholes,Gregory D.},
	doi = {10.1063/5.0007327},
	journal = {J. Chem. Phys.},
	number = {20},
	pages = {204101},
	title = {Generalization of the Hierarchical Equations of Motion Theory for Efficient Calculations with Arbitrary Correlation Functions},
	volume = {152},
	year = {2020}}

@article{olbrich2011atomistic,
  title={From atomistic modeling to excitation transfer and two-dimensional spectra of the FMO light-harvesting complex},
  author={Olbrich, Carsten and Jansen, Thomas LC and Liebers, J{\"o}rg and Aghtar, Mortaza and Stru{\"u}mpfer, Johan and Schulten, Klaus and Knoester, Jasper and Kleinekath{\"o}fer, Ulrich},
  journal={J. Phys. Chem. B},
  volume={115},
  number={26},
  pages={8609--8621},
  year={2011}
}

@article{camara2003structure,
  title={The structure of the FMO protein from Chlorobium tepidum at 2.2 {\AA} resolution},
  author={Camara-Artigas, Ana and Blankenship, Robert E and Allen, James P},
  journal={Photosynth. Res.},
  volume={75},
  pages={49--55},
  year={2003}
}

@article{schmidt2011eighth,
  title={The eighth bacteriochlorophyll completes the excitation energy funnel in the FMO protein},
  author={Schmidt am Busch, Marcel and Muh, Frank and El-Amine Madjet, Mohamed and Renger, Thomas},
  journal={J. Phys. Chem. Lett.},
  volume={2},
  number={2},
  pages={93--98},
  year={2011}
}

@article{tronrud2009structural,
  title={The structural basis for the difference in absorbance spectra for the FMO antenna protein from various green sulfur bacteria},
  author={Tronrud, Dale E and Wen, Jianzhong and Gay, Leslie and Blankenship, Robert E},
  journal={Photosynth. Res.},
  volume={100},
  pages={79--87},
  year={2009}
}

@article{milder2010revisiting,
  title={Revisiting the optical properties of the FMO protein},
  author={Milder, Maaike TW and Br{\"u}ggemann, Ben and van Grondelle, Rienk and Herek, Jennifer L},
  journal={Photosynth. Res.},
  volume={104},
  pages={257--274},
  year={2010}
}

@article{chandrasekaran2015influence,
  title={Influence of force fields and quantum chemistry approach on spectral densities of BChl a in solution and in FMO proteins},
  author={Chandrasekaran, Suryanarayanan and Aghtar, Mortaza and Valleau, St{\'e}phanie and Aspuru-Guzik, Al{\'a}n and Kleinekath{\"o}fer, Ulrich},
  journal={J. Phys. Chem. B},
  volume={119},
  number={31},
  pages={9995--10004},
  year={2015}
}

@article{beck2000multiconfiguration,
  title={The multiconfiguration time-dependent Hartree (MCTDH) method: a highly efficient algorithm for propagating wavepackets},
  author={Beck, Michael H and J{\"a}ckle, Andreas and Worth, Graham A and Meyer, H-D},
  journal={Phys. Rep.},
  volume={324},
  number={1},
  pages={1--105},
  year={2000}
}

@article{hase2020designing,
  title={Designing and understanding light-harvesting devices with machine learning},
  author={H{\"a}se, Florian and Roch, Lo{\"\i}c M and Friederich, Pascal and Aspuru-Guzik, Al{\'a}n},
  journal={Nat. Commun.},
  volume={11},
  number={1},
  pages={4587},
  year={2020}
}

@article{chen2022simulation,
  title={Simulation of absorption spectra of molecular aggregates: A hierarchy of stochastic pure state approach},
  author={Chen, Lipeng and Bennett, Doran IG and Eisfeld, Alexander},
  journal={J. Chem. Phys.},
pages={124109},
  volume={156},
  number={12},
  year={2022}
}

@article{varvelo2021formally,
  title={Formally exact simulations of mesoscale exciton dynamics in molecular materials},
  author={Varvelo, Leonel and Lynd, Jacob K and Bennett, Doran IG},
  journal={Chem. Sci.},
  volume={12},
  number={28},
  pages={9704--9711},
  year={2021}
}

@article{wen2009membrane,
  title={Membrane orientation of the FMO antenna protein from Chlorobaculum tepidum as determined by mass spectrometry-based footprinting},
  author={Wen, Jianzhong and Zhang, Hao and Gross, Michael L and Blankenship, Robert E},
  journal={Proc. Natl. Acad. Sci.},
  volume={106},
  number={15},
  pages={6134--6139},
  year={2009}
}

@article{higashi2016quantitative,
  title={Quantitative evaluation of site energies and their fluctuations of pigments in the Fenna--Matthews--Olson complex with an efficient method for generating a potential energy surface},
  author={Higashi, Masahiro and Saito, Shinji},
  journal={J. Chem. Theory Comput.},
  volume={12},
  number={8},
  pages={4128--4137},
  year={2016}
}

@article{kim2020toward,
  title={Toward monitoring the dissipative vibrational energy flows in open quantum systems by mixed quantum--classical simulations},
  author={Kim, Chang Woo and Rhee, Young Min},
  journal={J. Chem. Phys.},
  volume={152},
  number={24},
pages= {244109},
  year={2020}
}

@article{kim2014improving,
	title = {Improving long time behavior of Poisson bracket mapping equation: A non-Hamiltonian approach},
	volume = {140},
	issn = {0021-9606},
	url = {https://doi.org/10.1063/1.4874268},
	doi = {10.1063/1.4874268},
	pages = {184106},
	number = {18},
	journal = {J. Chem. Phys.},
	author = {Kim, Hyun Woo and Rhee, Young Min},
	year = {2014}
}

@article{sokolov2024non,
  title={Non-adiabatic molecular dynamics simulations provide new insights into the exciton transfer in the Fenna--Matthews--Olson complex},
  author={Sokolov, Monja and Hoffmann, David S and Dohmen, Philipp M and Kr{\"a}mer, Mila and H{\"o}fener, Sebastian and Kleinekath{\"o}fer, Ulrich and Elstner, Marcus},
  journal={Phys. Chem. Chem. Phys.},
  volume={26},
  number={28},
  pages={19469--19496},
  year={2024}
}

@article{kim2012all,
  title={All-atom semiclassical dynamics study of quantum coherence in photosynthetic Fenna--Matthews--Olson complex},
  author={Kim, Hyun Woo and Kelly, Aaron and Park, Jae Woo and Rhee, Young Min},
  journal={J. Am. Chem. Soc.},
  volume={134},
  number={28},
  pages={11640--11651},
  year={2012}
}

@article{nalbach2015vibronically,
  title={Vibronically coherent speed-up of the excitation energy transfer in the Fenna-Matthews-Olson complex},
  author={Nalbach, P and Mujica-Martinez, CA and Thorwart, M},
  journal={Phys. Rev. E},
  volume={91},
  number={2},
  pages={022706},
  year={2015},
  publisher={APS}
}

\end{document}